%% file: main.tex
\documentclass[a4paper,11pt]{article}
\usepackage{graphicx}
\usepackage{fullpage}
\pdfoutput=1 

\usepackage{mystyle} 

\usepackage{booktabs}
\usepackage{subcaption}
\usepackage{framed}
\usepackage{multirow}
\usepackage{array}
\usepackage{lscape}
\title{\sffamily Enhanced $thj$ signal at the LHC with $h\rightarrow \gamma\gamma$ decay and  $\mathcal{CP}$-violating top-Higgs coupling}

\author[]{Jason Yue
}

\affiliation[]{ARC Centre of Excellence for Particle Physics at the Terascale, School of Physics, The University of Sydney, NSW 2006, Australia}

\emailAdd{j.yue@physics.usyd.edu.au}

\abstract{
We study the observability of non-standard top Yukawa couplings in the $pp\rightarrow t(\rightarrow \ell \nu_\ell b ) h(\rightarrow \gamma\gamma)j$ channel at 14 TeV high-luminosity LHC (HL-LHC). The small diphoton branching ratio is enhanced when the $\mathcal{CP}$-violating phase, $\xi$, of the top-Higgs interaction is non-zero. When the modulus of the top-Higgs interaction assumes the SM value, $y_t=y_t^{SM}$, we find that the signal significance reaches $2.7\sigma$ $( 7.7\sigma)$ when $\xi= 0.25\pi$ $(0.5\pi) $. Furthermore, the different couplings modify the polarisation of the top quark, and can subsequently be distinguished via asymmetries in spin correlations with the final state leptons.  The diphoton decay mode is found to be  significantly more promising than the previously considered $pp\rightarrow t(\rightarrow \ell \nu_\ell b ) h(\rightarrow b\overline{b})j$ channel.}

\begin{document} 
\maketitle
%
%
%

\section{Introduction}
Since the discovery of the Higgs resonance \cite{Aad:2012, Chatrchyan:2012}, the measured properties \cite{ATLAS:2014_coupling,CMS:2014ega} of this boson have been consistent with a minimal Higgs sector. In particular, there has been compelling evidence that it carries a spin quantum number of zero \cite{Aad:2013xqa,Chatrchyan:2012jja,Chatrchyan:2013iaa,Chatrchyan:2013mxa} (for a discussion on a generic spin-2 impostor, see \cite{Kobakhidze:2013cna}).  The next priority is to establish the $\mathcal{CP}$-properties of this Higgs boson,  as this is important for pinning down new physics together with the associated cosmological implications. 
Currently, pure scalar couplings are favoured over pseudoscalar couplings to the electroweak gauge bosons \cite{Aad:2013xqa,Chatrchyan:2013mxa}. However, a non-minimal Yukawa sector containing substantial pseudoscalar admixture is not yet excluded (cf. \cite{Kobakhidze:2014gqa}) 
and requires further investigation.

The top-Higgs Yukawa coupling is the largest in the Standard Model (SM) and therefore plays an important role in electroweak symmetry breaking, notably in the context of Higgs vacuum stability (see e.g. \cite{Choudhury:2012tk,Buttazzo:2013uya,Spencer-Smith:2014woa}). Also, a $\mathcal{CP}$-violating top-Higgs sector may provide additional sources of $\mathcal{CP}$-violation that may have implications for the electroweak phase transition and baryogenesis \cite{Shu:2013uua,Zhang:1994fb}.   
 %
 Direct constraints on non-standard top Yukawa couplings must come from observation of processes where it enters at the tree-level, with the leading contribution coming from Higgs production associated with a top pair,  $pp\rightarrow t\overline{t}h$ (see e.g. \cite{Kunszt:1984ri, RichterWas:1999sa,Belyaev:2002ua,Maltoni:2002jr,Lafaye:2009vr,Biswas:2014hwa}). 
Specifically, the prospect for the LHC to distinguish the scalar and pseudoscalar components of the top Yukawa coupling in this channel have been considered  \cite{Gunion:1996xu, Gunion:1998hm, Bernreuther:1998qv, Han:2009ra, Frederix:2011zi,  Ellis:2013yxa, Demartin:2014fia,Khatibi:2014bsa}. Given the relatively small $pp\rightarrow t\overline t h$ cross section ($\sim$ 130 fb at 8 TeV \cite{Heinemeyer:2013tqa}), 
current luminosity and analysis are not yet sensitive enough to observe such signal. Nonetheless, an upper limit on the signal strength, $\mu_{t\overline{t}h}=\sigma_{t\overline{t}h}/\sigma^{SM}_{t\overline{t}h}<3.9$  has been set by the  ATLAS  collaboration at  95\% C.L. limit, using a combination of the $h \rightarrow b \overline{b}$ and $h \rightarrow \gamma\gamma$ channels \cite{atlas:tth_gaga}. A corresponding limit of $\mu_{t\overline{t}h}\in [0.9, 3.5] $ is established by the CMS collaboration using all search channels \cite{Khachatryan:2014qaa}.
 
This work focuses on the $pp\rightarrow thj$ production \cite{DiazCruz:1991cs,Stirling:1992fx,Ballestrero:1992bk,Bordes:1992jy,Tait:2000sh,Maltoni:2001hu,Campbell:2009ss,Cao:2007ea,Barger:2009ky}, despite its smaller SM cross section (18  fb at 8 TeV \cite{Farina:2012xp,Frederix:2009yq,Hirschi:2011pa,Frederix:2011ss}). 
The justification is that the interference between the $thj$ Feynman diagrams with $t\overline t h$ and $WWh$ couplings  \cite{Maltoni:2001hu, Farina:2012xp, Agrawal:2012ga,Biswas:2012bd,Ellis:2013yxa, Englert:2014pja, Kobakhidze:2014gqa,Barger:2009ky,Biswas:2013xva,Chang:2014rfa} makes it more sensitivity to the $\mathcal{CP}$-violating phase, $\xi$, of the top Yukawa coupling, in addition to its modulus, $y_t$. Specifically, increased $|\xi|$ values reduce the $pp\rightarrow t\overline{t}h$ cross section \cite{Ellis:2013yxa}, but enhance the $pp\rightarrow thj$ cross section    \cite{Kobakhidze:2014gqa,Chang:2014rfa}. 
{ 
 Due to this interference, the extraction of the top Yukawa coupling will require a sufficiently precise measurement of the $hWW$ coupling, which is expected to reach the $2$-$5\%$ level at 3000 fb$^{-1}$ at the LHC \cite{Dawson:2013bba}.}  A further advantage of the single top channel is that the resulting $t$-quark is polarised by the left-handed weak interaction through a $t$-channel virtual $W$ boson, but not in the top pair channel. 
As the $t$-quark decays before it hadronises, spin information is inherited by its decay products \cite{Allanach:2006fy,Godbole:2006tq,Shelton:2008nq,Godbole:2009dp,Krohn:2009wm,Godbole:2010kr,Falkowski:2011zr}. Top quark polarisation induced by the $\mathcal{CP}$-violating  top-Yukawa couplings will therefore manifest in the spin correlation of the top decay products  (see e.g. \cite{BhupalDev:2007is,Godbole:2006eb,Bernreuther:1997gs,Bernreuther:1998qv,Ellis:2013yxa,Englert:2014pja,Korchin:2014kha,Beneke:2000hk}). Polarisation in  single top processes  \cite{Aguilar-Saavedra:2014eqa,Prasath:2014mfa,Taghavi:2013wy,Godbole:2011vw,Rindani:2011gt,Rindani:2011pk,Huitu:2010ad,Arai:2010ci}\footnote{Information on the $\mathcal{CP}$-violating $t\overline{t}h$ couplings can also be inferred from angular correlation in $pp\rightarrow h+2j $ \cite{Klamke:2007cu,Hankele:2006ja,Dolan:2014upa}.} have specially been proposed for probing non-trivial chiral structures in top couplings. 

The $h\rightarrow b\overline{b}$ decay mode in $thj$ production is found to be a very challenging channel for probing $\mathcal{CP}$-violation in the top-Higgs sector  due to large QCD backgrounds  \cite{Kobakhidze:2014gqa, CMS:2014zqa}. It is therefore important to explore  
$h\rightarrow \gamma \gamma$ decay, which offers  much more manageable backgrounds. The key feature of this decay mode, unlike $h\rightarrow b \overline{b}$, is that it is mediated via $t$- and $W$-loops, giving it sensitivity to the $t\overline{t}h$
couplings. With the diphoton branching ratio increasing with $|\xi|$, it is conceivable that $h\rightarrow \gamma\gamma$ will be the more favourable channel to probe the $\mathcal{CP}$-phase. However, these advantages are possibly offset by the small SM diphoton branching ratio, and given that searches for $thj$ signal have already began by the CMS collaboration  \cite{CMS:2014oca}  in the $h\rightarrow \gamma\gamma$ channel, a detector-level analysis for this channel is necessary. We remark that the photon polarisations of the diphoton decay \cite{Bishara:2013vya,Asakawa:2000jy} may provide information on$\xi$, but this will not be considered in this work.

This paper will be organised as follows:  in Sec.~\ref{sec:lagrangian}, the enhancement of the diphoton branching ratio through $\mathcal{CP}$-violating top-Higgs couplings, and its consistency with current Higgs data is discussed; Sec.~\ref{sec:signal} will be concerned with the observability of the $thj$ signal with scalar ($\xi=0$), pseudoscalar ($\xi=0.5\pi$) and mixed ($\xi=0.25\pi$) top-Higgs interactions at 14 TeV high luminosity LHC (HL-LHC); the use of lepton spin correlation and asymmetries to distinguish the various phases is studied in Sec.~\ref{sec:pol}; the results are summarised in the conclusion (Sec.~\ref{sec:conclusion}). 
\section{$\mathcal{CP}$-violating Top-Higgs Sector and Enhanced $h\gamma\gamma$ Decay Rate }\label{sec:lagrangian}
In this study, we investigate the $\mathcal{CP}$-violating top Yukawa couplings using the phenomenological Lagrangian: 
 \begin{align}
 \label{eqn:pheno_yt}
   \mathcal{L}_{\text{pheno} }&\supset-\frac{y_t}{\sqrt{2}} \overline{t}(\cos\xi +i\gamma^5 \sin \xi) t h,
\end{align}
where $t$ and $h$ are respectively the physical top quark and Higgs boson in the mass basis, $y_t \in \mathbb{R}$ parameterises the magnitude of the $t\overline{t}h$ interactions and $\xi \in (-{\pi},{\pi}]$ is the $\mathcal{CP}$-violating phase. In the SM limit, $y_t$ takes the value $y_t^{SM}:={\sqrt{2}m_t}/{v}$ and $\xi = 0$, where $v\approx 246$ GeV is the vacuum expectation value of the Higgs field. Such a non-standard top-Higgs sector may arise from various beyond SM models \cite{Han:2013sea,McKeen:2012av,Celis:2013rcs,Kobakhidze:2012wb,Bernreuther:1998qv,Accomando:2006ga}. 
The framework used in this work will be that of an effective field theory whereby phenomenological predictions can be made without adhering to specific models. The Lagrangian in Eqn.~\ref{eqn:pheno_yt} may originate from that of an 
 effective one comprising gauge-invariant operators\cite{AguilarSaavedra:2008zc,AguilarSaavedra:2009mx,Zhang:2012cd,Belusca-Maito:2014dpa,He:2013tia}:  
 \begin{align}
   \mathcal{L}_{\text{dim} \leq 6} &\supset -\bigg( \alpha + \beta \frac{ H^\dagger H}{\Lambda^2}\bigg) H Q_L^\dagger t_R + h.c.,
\end{align}
where $\alpha,\beta\in\mathbb{C}$ are dimensionless parameters and $\Lambda$ the new physics scale.  After symmetry breaking with $H=(0,v+h/\sqrt{2})^T$, it may be identified that  $y_t^{SM}=\alpha+\beta v^2/\Lambda^2$. The phase $\xi$ may take the full range $(-\pi,\pi]$, given that new physics enters at the TeV scale $(\Lambda \sim 10v)$ and $|\beta|\sim 1$ \cite{Harnik:2013aja}.

An immediate consequence of non-standard top-Yukawa couplings are deviations of $gg\rightarrow h$ production and  $h\rightarrow\gamma\gamma$ decay rates from the SM. This has been considered in our previous work \cite{Kobakhidze:2014gqa}, where the current Higgs data was used to exclude values of  $|\xi|>0.6\pi$ at 95\% C.L.  (see also \cite{Ellis:2013yxa,Nishiwaki:2013cma,Belusca-Maito:2014dpa,Djouadi:2013qya,Bhattacharyya:2012tj}). The corresponding 95\% C.L. limit allowed for $y_t /y_t^{SM}$ is $[0.7,1.2]$ when $\xi =0$, but decreases with $\xi$ such that it becomes $[0.4,0.6]$ for $\xi=0.5\pi$. Strong bounds on $\mathcal{CP}$-violating effects also come from low energy probes \cite{Atwood:2000tu,Cheung:2014oaa,Brod:2013cka} such as electric dipole moments. However, such bounds on the pseudoscalar coupling are not considered in this work as they depend on the light fermion Yukawa couplings which are practically unobservable at the LHC. In the subsequent parts of this work, only the top Yukawa sector will be modified, but $y_t=y_t^{SM}$ will be assumed to focus on the effects of $\xi$. 

We will now focus on the influence of non-standard top Yukawa coupling in $h\rightarrow \gamma\gamma$, as it is the relevant decay mode for the phenomenological study in Sec.~\ref{sec:signal} and Sec.~\ref{sec:pol}.  The associated effective operator for $h\gamma\gamma$ interactions can be written as  (see e.g. \cite{Korchin:2013ifa,Shifman:1979eb,Voloshin:2012tv}):   
   \begin{align}
\mathcal{L}_{h\gamma\gamma} = \frac{\alpha}{8\pi v}\bigg( c_\gamma F_{\mu\nu}F^{\mu\nu} - \tilde{c}_\gamma F_{\mu\nu}\tilde{F}^{\mu\nu}\bigg)h,
         \end{align}
         where $\alpha=e^2/4\pi $ is the fine structure constant, $F_{\mu\nu}$ is the standard field strength for photon fields and $\tilde{F}^{\mu\nu}:=\frac{1}{2}\epsilon^{\mu\nu\rho\sigma}F_{\rho\sigma}$ is its dual. Given that the $\mathcal{CP}$-even and $\mathcal{CP}$-odd parts do not interfere, the effective scalar ($c_\gamma$) and  pseudoscalar ($\tilde{c}_\gamma$) coupling constants are obtained from the corresponding scalar and pseudoscalar decay rates, $ \Gamma_{S,P}(h\rightarrow \gamma\gamma)$ (cf. Appendix~\ref{app:funcs}). At one loop order, the scalar part is dominated by $t$-quark and $W$-boson contributions whilst the latter is absent for the pseudoscalar part. Accordingly, the coupling constants are parameterised in terms of the top Yukawa couplings as follows:
 \begin{eqnarray}
c_\gamma &\approx& -8.32  +1.83 {y_t}\cos\xi/{y_t^{SM}},\\
\tilde{c}_\gamma  &\approx & 2.79{y_t}\sin\xi/{y_t^{SM}}.
\end{eqnarray}
Assuming that no other new physics processes contribute to the $h\gamma\gamma$ loop, the total diphoton decay rate resulting from the modified top-Higgs sector is then parameterised as:
\begin{eqnarray}\label{eqn:branch}
\Gamma(h\rightarrow\gamma\gamma)\approx\frac{m_h^3\alpha^2}{256\pi^3v^2} \bigg[\left(-8.32  +\frac{1.83 {y_t}\cos\xi}{y_t^{SM}}\right)^2  + \left( \frac{2.79{y_t}\sin\xi}{y_t^{SM}}\right)^2\bigg].
\end{eqnarray}   
It should be noted that the partial cancellation between the $W$-loop and $t$-loop contributions to the scalar component diminishes with increasing $\xi$.  Since the pseudoscalar $t$-loop factor is larger than that of the scalar, and the interaction with different parities do not interfere,  the decay rate will be maximally enhanced for $\xi=0.5\pi$.
                   
\section{Observability at 14 TeV HL-LHC}\label{sec:signal}
The sensitivity of the Higgs associated single top production channel to the $\mathcal{CP}$-violating top-Higgs couplings at 14 TeV HL-LHC was investigated through Monte Carlo simulations. This was carried out in the diphoton decay of the Higgs and semileptonic decay of the top:
\begin{equation}
pp\rightarrow{t}(\rightarrow \ell ^+{\nu}_\ell {b}) h(\rightarrow \gamma\gamma)j,
\end{equation}
with $j$ denoting the light jets and $\ell = e,\mu$.  Three phases $\xi=0$, $0.25\pi$ and $0.5\pi$ were studied as benchmark points, with $y_t$ assuming the SM value $y_t^{SM}$, as justified in Sec.~\ref{sec:lagrangian}. {  The observability of the signal process will not be affected the inverting the signs of the $\xi$ values, as the production cross section (cf. \cite{Kobakhidze:2014gqa}) and decay rate (cf. Eqn.~\ref{eqn:branch}) are not sensitive to this sign difference. We therefore  examine only the positive phases.}
The small branching ratio $Br_{SM} (h\rightarrow \gamma\gamma)=2.28\times 10^{-3}$ is compensated by excellent resolution on invariant diphoton mass, $m_{\gamma\gamma}$ at ATLAS and CMS. Continuous QCD backgrounds can therefore be efficiently suppressed by a narrow mass window cut on the reconstructed Higgs mass, $m_h$. Furthermore, it could be seen from Eqn.~\ref{eqn:branch} that the decay rate may be enhanced to $\sim 1.3 \ (1.8)$ that of the SM for $\xi=0.25\pi \ (0.5\pi) $. The observability of such signal may be further improved, given that the $\mathcal{CP}$-violating phase also enhances the $thj$ production cross section \cite{Kobakhidze:2014gqa}.  The semileptonic decay mode of the top quark is chosen because the charged lepton is maximally correlated with the top spin \cite{Jezabek:1988ja} and will be used to measure the forward-backward asymmetry in Sec.~\ref{sec:pol}.

The dominant backgrounds to the signal process in order of contribution are as follows:
\begin{itemize}
  \item[(B1)] $t(\rightarrow \ell ^+{\nu} b)j \gamma\gamma$ --- this irreducible background has the same final state as the signal. However it is non-resonant and is expected to be efficiently suppressed through a window cut on $m_{\gamma\gamma}$.
  \item[(B2)] $t(\rightarrow \ell ^+ {\nu}{b})\overline{t}(\rightarrow \overline{b} j j) \gamma\gamma$ ---  one of the jets in the hadronically decaying top is misidentified, and the other two are missed in the detector. The analysis will include $h\rightarrow \gamma \gamma $ contributions for each $\xi$.
  \item[(B3)] $W^+(\rightarrow \ell ^+ \overline{\nu} )\gamma\gamma jj$ --- one jet is mis-tagged as a $b$-jet and the other missed in the detector. Again, the photon pair may result from Higgs decay. This background is excluded in the analysis since it was demonstrated in \cite{Chang:2014rfa} and \cite{Wu:2014dba} to be at least an order of magnitude lower than (B1) and (B2) after a window cut on $m_{\gamma\gamma}$. 
\end{itemize}

The effective Lagrangian in Eq.~\ref{eqn:pheno_yt} was implemented by \texttt{FeynRules} \cite{Alloul:2013bka} with SM parameters taken from \cite{Beringer:1900zz}. The signal and background matrix elements were generated by \texttt{MadGraph 5} package \cite{Alwall:2014hca}  with default parton level cuts, and convolved with the CTEQ6L parton distribution function \cite{Pumplin:2002vw} using default dynamical renormalisation ($\mu_R$) and factorisation ($\mu_F$) scale. Parton showering was subsequently performed by  \texttt{Pythia} \cite{Sjostrand:2007gs} and jets were clustered via anti-$k_t$ algorithm \cite{Cacciari:2008gp} with a cone radius of $\Delta R = 0.7$. Detector simulation was carried out by \texttt{Delphes} \cite{deFavereau:2013fsa} where the (mis-)tagging efficiencies and fake rates assume their default values.  

\begin{figure}[h!]
\centering
\includegraphics[width=.45\textwidth,clip=true,trim=5mm 5mm 10mm 16mm]{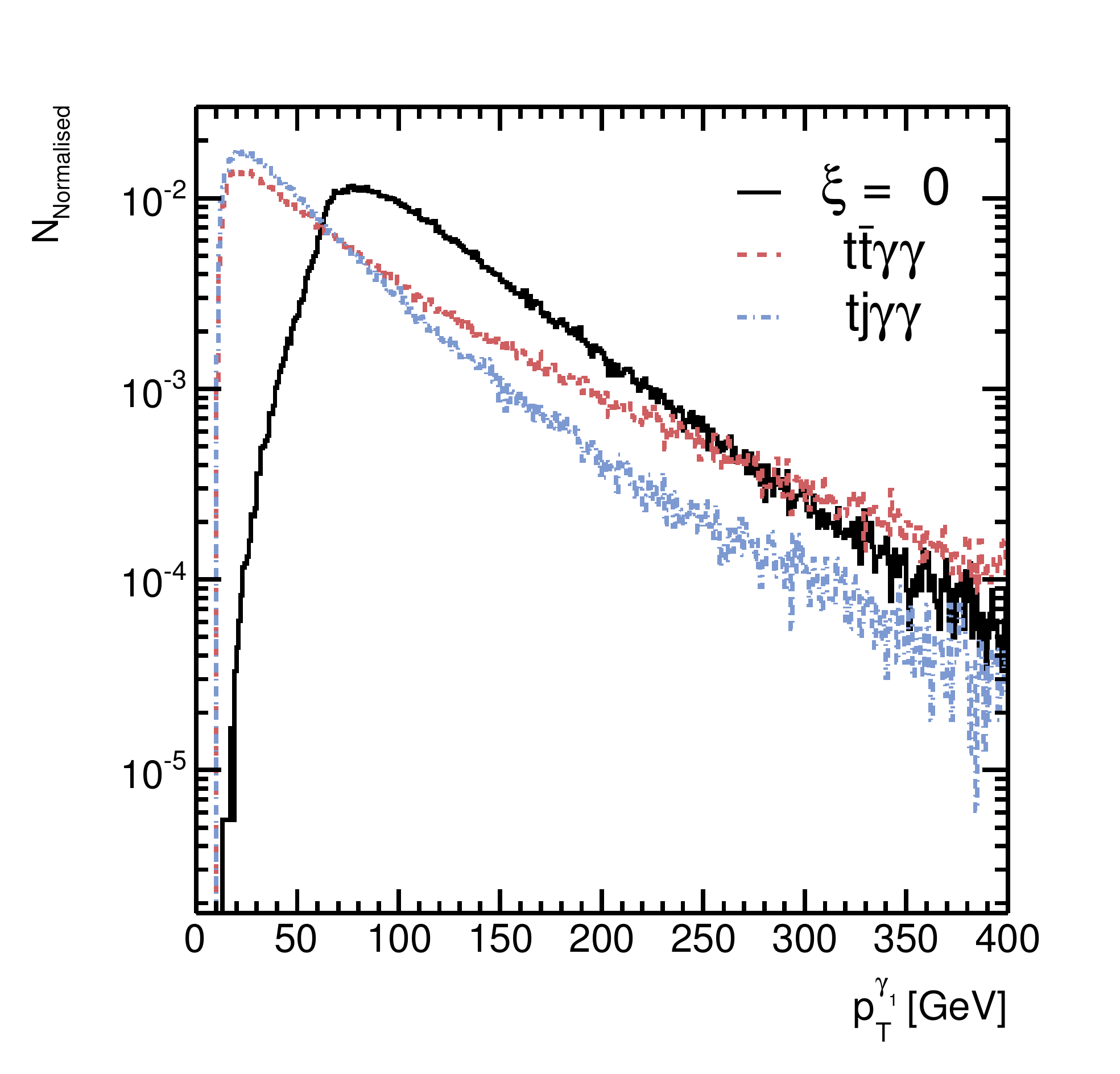}
\includegraphics[width=.45\textwidth,clip=true,trim=5mm 5mm 10mm 16mm]{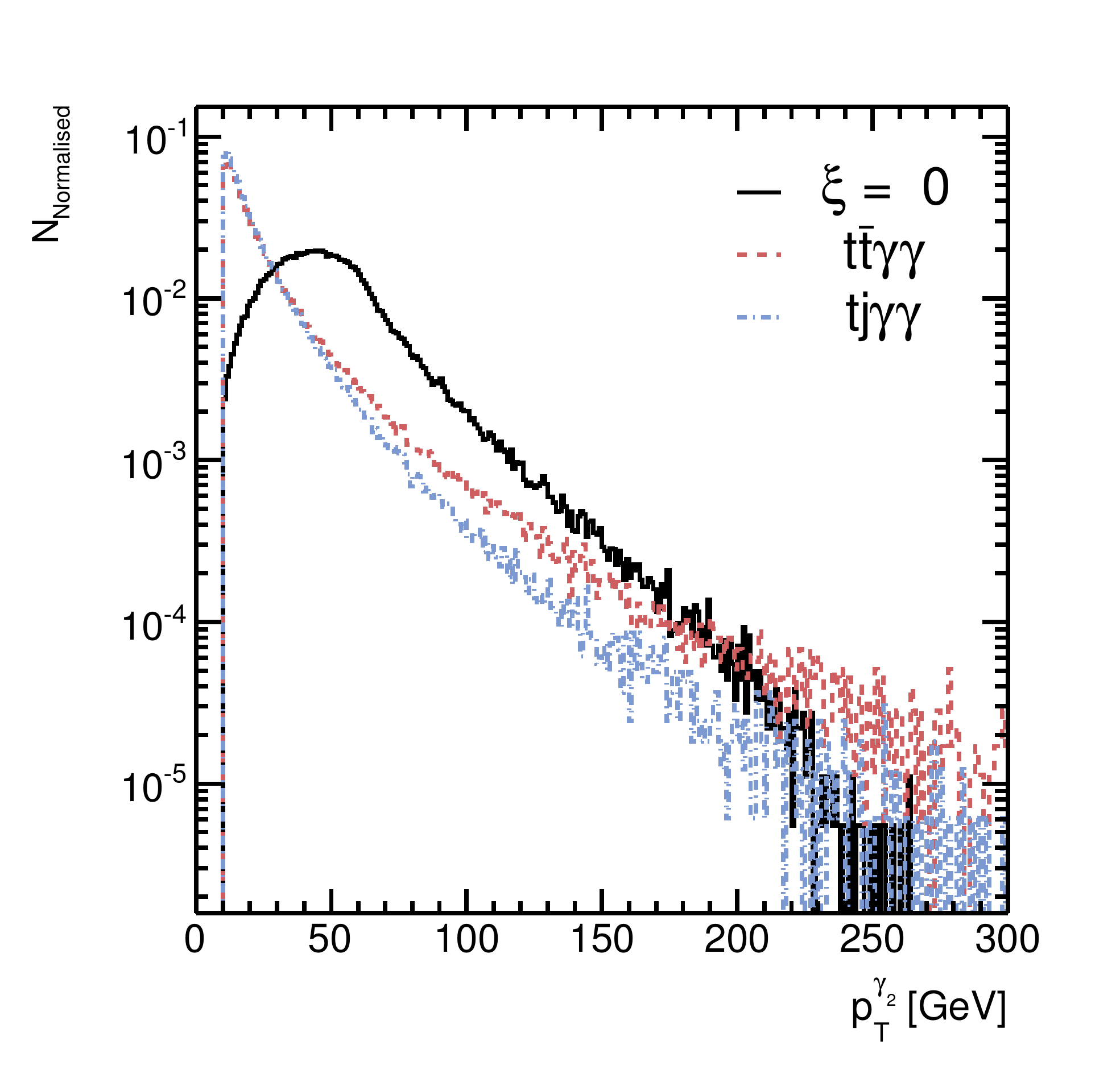}
\caption{$p_T$ of the leading jet (left) and the subleading jet (right).}
\label{fig:ga_pt}
\end{figure}

The signal analysis (cf. Tab.~\ref{tab:14Tev_cutflow}) begins with the basic selection criteria (C1) on transverse momenta and rapidities, based on the trigger capabilities and detector coverage at the LHC. In Fig.~\ref{fig:ga_pt}, it is evident that the $p_T$ spectra of the leading  ($\gamma_1$) and subleading  ($\gamma_2$)  photons in the SM signal are more energetic than that of the corresponding backgrounds. The $thj$ signals with $\xi=0.25\pi$ and $0.5\pi$ also exhibit the same behaviour. Furthermore,  the resonant production of diphoton pairs in the signals leads to a peak near $m_h/2$,  allowing them to be separated from the non-resonant diphoton pairs in $t\overline{t}\gamma\gamma$ and $tj\gamma\gamma$ through the $p_T^{\gamma_1} > 50$ GeV and $p_T^{\gamma_2} > 25$ GeV cut (C2) in Tab.~\ref{tab:14Tev_cutflow}.

\begin{centering}
\begin{table}[h]
\fontsize{7pt}{8.4pt}\selectfont
\begin{center}
{\renewcommand{\arraystretch}{1.25}
\newcolumntype{C}[1]{>{\centering\let\newline\\\arraybackslash\hspace{0pt}}m{#1}} 
\begin{tabular}{ ccc| ccc |ccc |  c}
\hline 
&\multicolumn{2}{c|}{\multirow{3}{*}{Cuts}}&\multicolumn{7}{c}{$\sigma$ [$10^{-3} $ fb] } \\\cline{4-10}
&&&\multicolumn{3}{c|}{$t(\rightarrow\ell\nu_\ell b)h(\rightarrow \gamma \gamma )j$}&\multicolumn{3}{c|}{$t\overline{t}\gamma\gamma$ }&{\multirow{2}{*}{$tj \gamma\gamma$}} \\\cline{4-9}
&&&{$\xi=0$} & {$\xi=0.25\pi$} &{$\xi=0.5\pi$}  &  {$\xi=0$} & {$\xi=0.25\pi$} &{$\xi=0.5\pi$}  &  \\\cline{1-10}
\multirow{5}{*}{(C1)} &$ \Delta R_{ij} > 0.4$ & $i,j=b,j,\ell,\gamma$
& \multirow{5}{*}{4.545}& \multirow{5}{*}{10.32}& \multirow{5}{*}{42.79}
& \multirow{5}{*}{145.0}& \multirow{5}{*}{145.8}& \multirow{5}{*}{144.4}& \multirow{5}{*}{299.4}\\
&   $p_{T}^b > 25 \ \text{GeV}, $     &$|\eta_b|<2.5$&&&&&&&\\
&     $p_{T}^\ell > 25 \ \text{GeV}$,    &$|\eta_\ell|<2.5$&&&&&&\\
&    $ p_{T}^j > 25 \ \text{GeV}$,    &$|\eta_j|<4.7$&&&&&&\\   
&    $ p_{T}^\gamma > 20 \ \text{GeV}$,    &$|\eta_\gamma|<2.5$&&&&&&&\\   
 \hline
 (C2)&      { $ p_T^{\gamma_1} > 50$ GeV},& { $ p_T^{\gamma_2} > 25$ GeV} &4.194 &9.599& 39.69
 &88.11&88.24&87.59&155.2\\ \hline
(C3)&      \multicolumn{2}{ c| }{ $M_{b\ell}<200\ \mathrm{GeV}$} &4.059&9.104&37.44&
64.05&64.10&63.68&151.3\\ \hline
(C4)& \multicolumn{2}{c|}{$ |M_{\gamma\gamma}-m_h|< 5 \ \text{GeV}$}&3.219&6.866&28.47&3.295&3.493&3.393&9.031\\\hline
\hline
\multicolumn{3}{c|}{$S/B$ }   &0.261 &0.548& 2.29\\\hline
\multicolumn{3}{c|}{$S/\sqrt{S+B}$ with 3000 fb$^{-1}$}   &1.41 &2.70& 7.71
 \\ \hline
\end{tabular}}
\caption{ Cut flow of the cross sections for the signals and backgrounds at 14 TeV LHC. 
The $h\rightarrow \gamma\gamma$ contributions to the $t\overline{t}\gamma\gamma$  background are included.
Conjugate processes are included here. \label{tab:14Tev_cutflow}}
\end{center}
\end{table}
\end{centering}


\begin{figure}[h!]
\centering
\includegraphics[width=.45\textwidth,clip=true,trim=5mm 8mm 10mm 16mm]{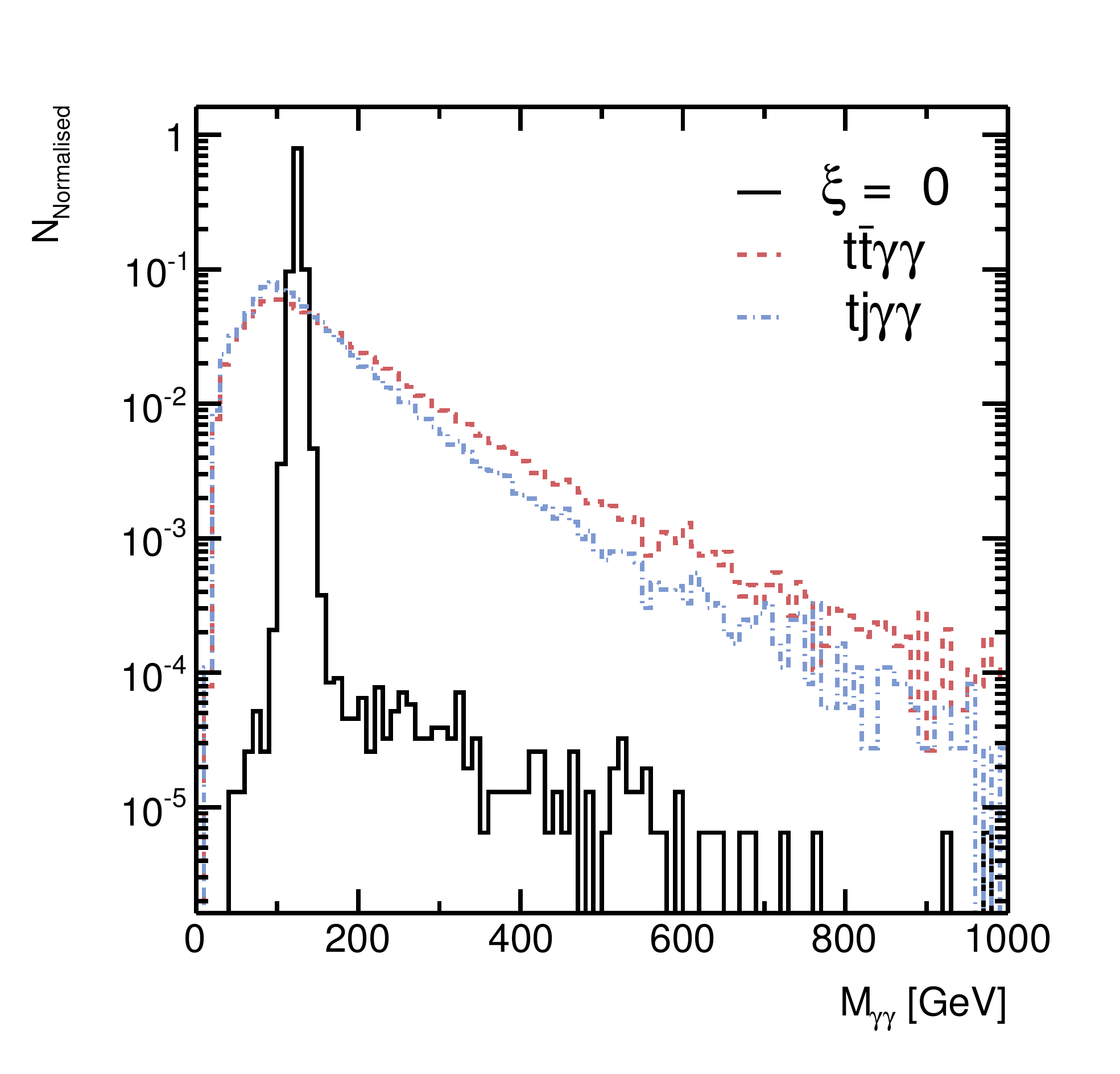}
\caption{Diphoton invariant mass for signal and backgrounds with $\xi=0$. The shapes are similar for $\xi=0.25\pi$ and $0.5\pi$.} 
\label{fig:diphoton}
\end{figure}

The cut (C3) on the invariant mass of the leading $b$-jet and lepton $\ell$ have been discussed in \cite{Kobakhidze:2014gqa, Chang:2014rfa,Wu:2014dba}. Given that both the $b$-jet and lepton $\ell$ should originate from the same $t$-quark, their invariant mass should be less than the top mass.  As the leading $b$-jet in $t\overline{t}\gamma\gamma$ could have came from either of the top quarks, it is not surprising to find that (C3) is more effective on this background when compared to the signals and $tj\gamma\gamma$. Lastly, Fig.~\ref{fig:diphoton} shows that the scalar signal has a relatively narrow diphoton invariant mass peak after (C3). It was also verified that $\xi =0.25\pi$ and $0.5\pi$ exhibit a similar distribution. The invariant mass window cut  $|m_{\gamma\gamma}-m_h|<5$ GeV (C4) is found to be the most effective, removing $\sim1/4$ of the signal events but the backgrounds by a factor of at least 16. Despite an increased $h\rightarrow \gamma\gamma$ contribution in $t\overline{t}\gamma\gamma$  for $\xi =0.25\pi$ and $0.5\pi$ due to the enhanced Higgs-diphoton decay rates, Tab.~\ref{tab:14Tev_cutflow} shows that the full $t\overline{t}\gamma\gamma$ cross-section remains relatively similar to the $\xi=0$ case after (C4). 

 The significance, $S/\sqrt{S+B}$ at the end of HL-LHC (3000 fb$^{-1}$), is expected reach 7.71$\sigma$ for $y_t=y_t^{SM}$ and $\xi=0.5\pi$. However, to   remain consistent with the current Higgs data, the modulus of the Yukawa coupling has to decrease to $y_t=0.6y_t^{SM}$ \cite{Kobakhidze:2014gqa} and the corresponding signal  significance is expected to drop to $\sim 4.75\sigma$. This is still significantly higher than that for the $h\rightarrow b\overline{b}$ mode. With an estimated uncertainty  of ${+3.1}/{-2.5}\%$ due to the parton distribution function \cite{CMS:2014oca},  the results are not expected to change significantly even with  a 10\% NLO correction on the production cross section \cite{Farina:2012xp}. However, the mixed and pure scalar couplings remain less optimistic for observation, and it is foreseeable that a combination of  the $\gamma\gamma$ and $b\overline{b}$ channels is required to achieve a high enough signal significance in these scenarios.

\section{Top Polarisation and Lepton Spin Correlation}\label{sec:pol}

The angular distribution of the lepton from a polarised top quark is given by \cite{Jezabek:1988ja}:

\begin{align}
\frac{1}{\Gamma} \frac{d\Gamma}{\cos \theta_\ell } =\frac{1}{2} (1+\mathcal{P}_t \kappa_\ell \cos \theta_\ell ),
\end{align}
where $\kappa_\ell$ is the lepton spin analysing power, $\theta_\ell$ is the angle between the lepton momenta and spin quantisation axis of the top, as measured in the rest frame of the $t$-quark, and $\mathcal{P}_t$ is the spin asymmetry. In this study, the top spin axis is taken to be the direction of the top quark in the laboratory frame. In order to reconstruct the top rest frame, the neutrino momentum was first determined from the on-shell condition of the $W$-boson \cite{Berger:2011xk,Xiao-ping:2014npa}:
\begin{align}
p_{\nu L}:=\frac{1}{2p^2_{\ell T}}\bigg(A_Wp_{\ell L}\pm E_\ell \sqrt{A_W^2-4\ p_{\ell T}^2 \not\!\!{E}_T^2}\bigg) \qquad \text{and} \qquad  p_{\nu T}:=\not\!\!E_T,
\end{align}
where $A_W:=m_W^2+2\mathbf{p}_T\cdot   { \not\!\!\mathbf{E}_T}$. The sign ambiguity is resolved via minimisation of: 
\begin{align}
\chi^2=\left(\frac{m_t-m_{\ell\nu_\ell b}}{\Gamma_t}\right)^2,
\end{align}
where $\Gamma_t$ is the SM top decay width. 
\begin{figure}[h!]
\centering
\includegraphics[width=.45\textwidth,clip=true,trim=7mm 5mm 10mm 16mm]{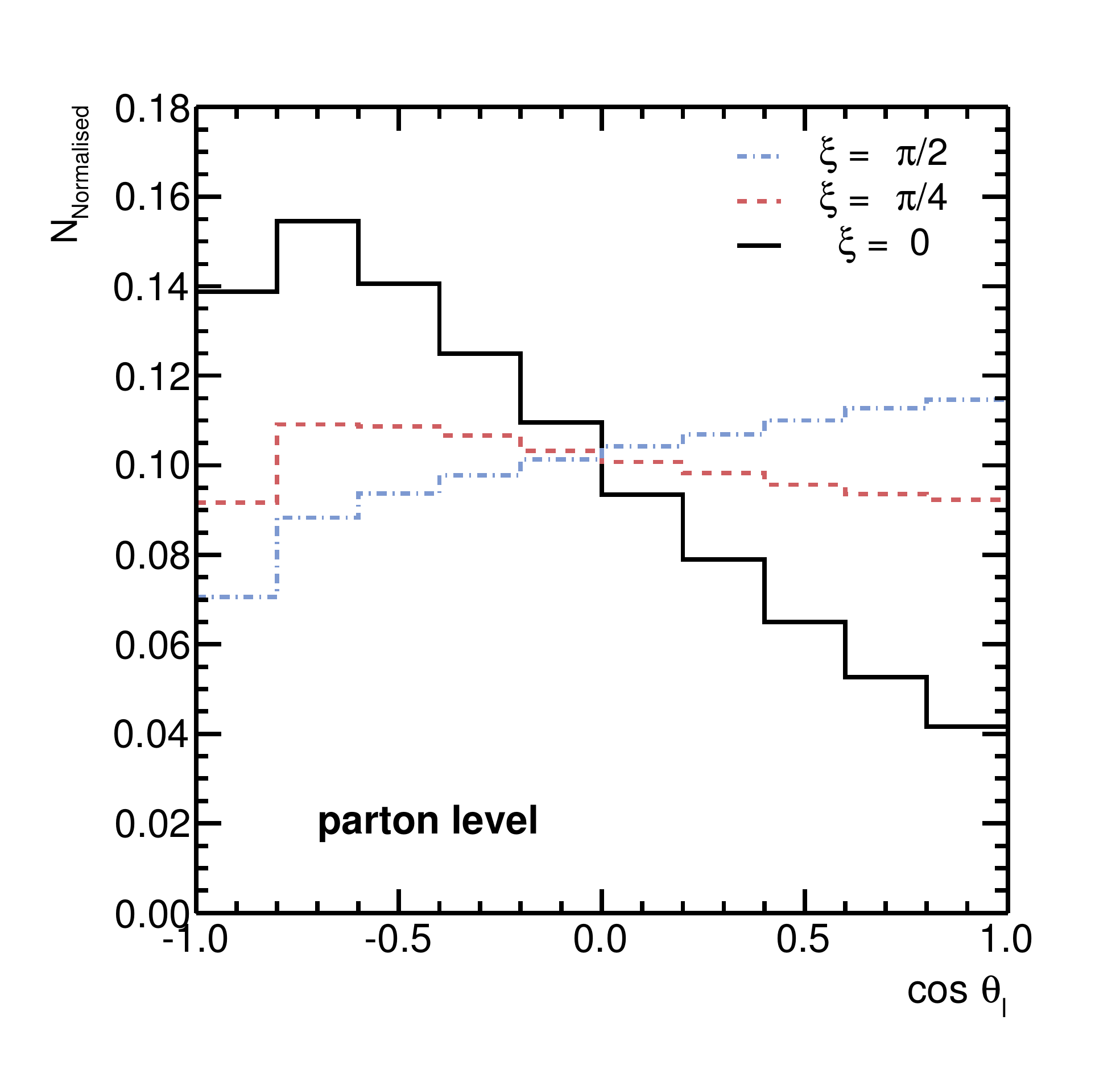}
\includegraphics[width=.45\textwidth,clip=true,trim=7mm 5mm 10mm 16mm]{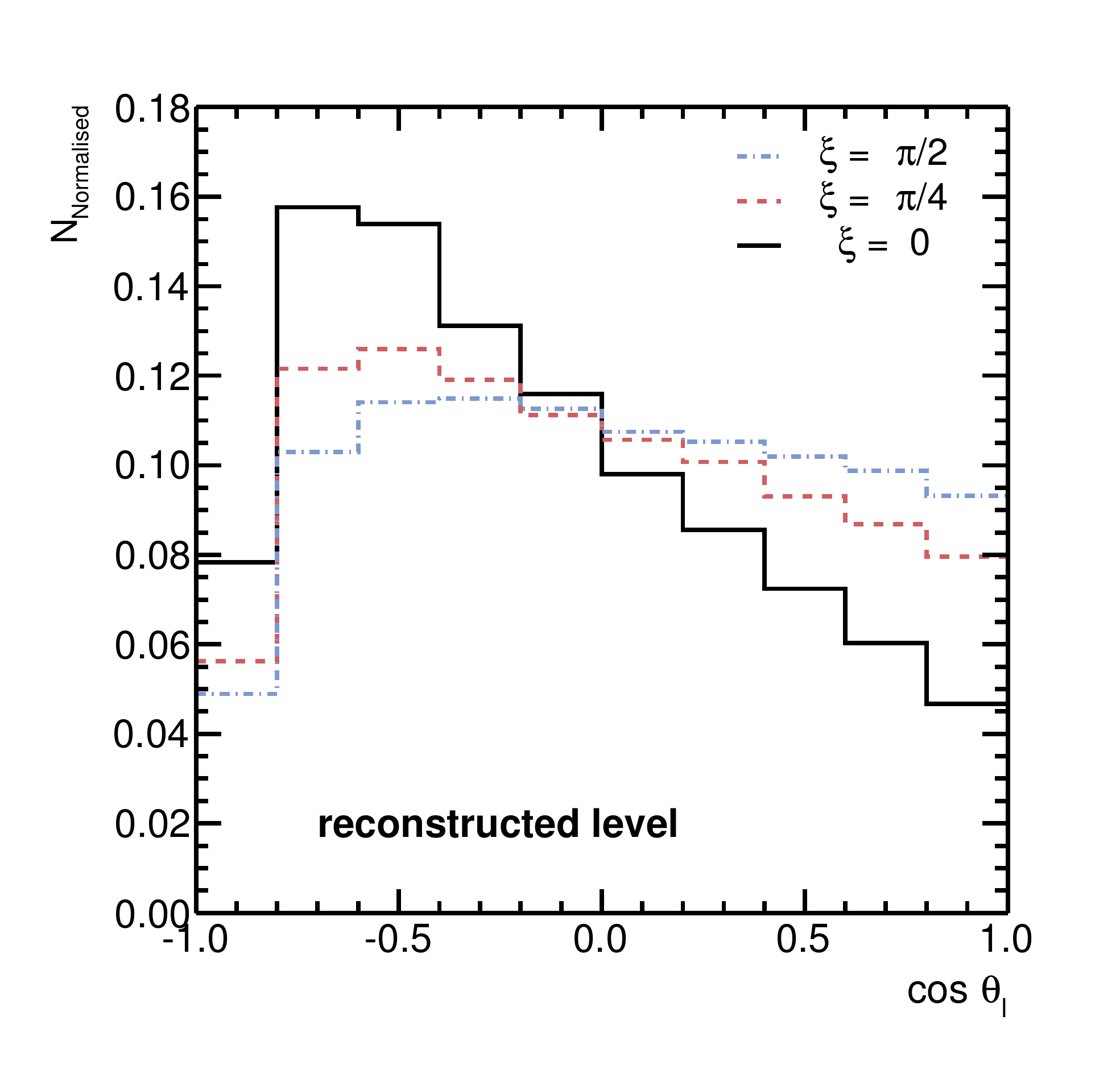}
\caption{The lepton angular correlation in the diphoton decay channel of $pp\rightarrow{t}(\rightarrow \ell^+ {\nu_\ell} {b}) h(\rightarrow \gamma\gamma)j$  at parton level (left), and reconstruction level (right) after the cuts.}
\label{fig:pol}
\end{figure}
\begin{table}[c h!]
\begin{center}
{\renewcommand{\arraystretch}{1.25}
\newcolumntype{C}[1]{>{\centering\let\newline\\\arraybackslash\hspace{0pt}}m{#1}}

\begin{tabular}{  C{1cm} C{4cm}  C{4cm}  C{2cm}C{2cm}}
\hline
 $\xi$& $\sigma(\cos\theta>0)$ [$10^{-4}$ fb] & $\sigma(\cos\theta<0)$  [$10^{-4}$ fb]& $\mathcal{A}^{\ell}_{FB}$ (\%) &Significance \\
\hline
{$0$} &4.413 &7.745 &-27.40&0.5234\\\hline
{$0.25\pi$} &12.05 &13.81 &-6.805&0.1895\\\hline
{$0.5\pi$} &54.21 &50.56 &3.484&0.1953\\\hline
\hline
\end{tabular}}
\caption{The reconstructed-level forward-backward asymmetry $\mathcal{A}^{\ell}_{FB}$ at 14 TeV LHC with 3000 fb$^{-1}$.}\label{tab:afb}
\end{center}
\end{table}
The lepton angular distributions for the $\mathcal{CP}$-phases $\xi = 0$, $0.25\pi$ and $0.5\pi$ are shown in Fig.~\ref{fig:pol} for the parton level and reconstructed level after cuts (C4). It is evident that in the SM ($\xi=0$), the preferential direction for the lepton momentum in the top rest frame is opposite to the top's boost. The pure pseudoscalar ($\xi=0.5\pi$) interaction changes the polarisation of the top through a $t\overline{t}h$ vertex, such that the lepton direction becomes positively correlated with the top's boost. As expected, the mixed interaction ($\xi=0.25\pi$) gives a slope that is intermediate between the pure scalar and pseudoscalar cases.  {  Again, the slope is not sensitive to the sign of $\xi$ \cite{Ellis:2013yxa}}. The difference between the slopes become less prominent in the reconstructed case, reflecting the simulated effects of parton showering, reconstruction efficiencies and detector resolution. The differences between the angular correlations are quantified in terms of the lepton forward-backward asymmetry:
\begin{align}\label{eqn:afb}
\mathcal{A}^\ell_{FB}:=\frac{\sigma(\cos\theta_\ell >0)-\sigma(\cos\theta_\ell <0)}{\sigma(\cos\theta_\ell >0)+\sigma(\cos\theta_\ell <0)},
\end{align}
and the significance \cite{Kao:1999kj,Liu:2010ay} by $\frac{|\Delta_\sigma|\mathcal{L}}{\sqrt{\sigma_T \mathcal{L}}}$,
with $\Delta_\sigma$ and $\sigma_T$ being the numerator and denominator of the right hand side of Eqn.~\ref{eqn:afb} respectively. From Tab.~\ref{tab:afb}, it is observed that  the top-Higgs interaction can be distinguished via $\mathcal{A}^\ell_{FB}$ with the SM case reaching a value of -27\% and a significance of 0.52 whilst that for the pseudoscalar case, 3.4\% with a significance of 0.20.

\section{Conclusion}\label{sec:conclusion}
In this work, we investigated the observability of the  $pp \rightarrow t(\rightarrow \ell^+ \nu_\ell b) h(\rightarrow \gamma\gamma)j$  at 14 TeV HL-LHC. The detector resolution on $m_{\gamma\gamma}$ allows effective suppression of QCD background via a  mass window cut, compensating for its small diphoton branching ratio. In addition, non-zero $\xi$ enhanced the $pp\rightarrow thj$ production cross section and the $h\rightarrow \gamma\gamma$ branching ratio. The $\gamma\gamma$ channel is found to give much better prospects than the $b\overline{b}$ channel to probe the  $\mathcal{CP}$-violating top-Higgs couplings. With  $y_t=y_t^{SM}$, the expected signal significances are 1.4$\sigma$, 2.7$\sigma$ and 7.7$\sigma$  for scalar $(\xi=0)$, mixed ($\xi=0.25\pi$) and pseudoscalar ($\xi=0.5\pi$) interactions respectively. Furthermore, the diphoton channel led to measurable differences in lepton spin correlation by modifying the $t$-quark polarisation and can be distinguished via the forward-backward asymmetries. 

\acknowledgments

I would like to thank Archil Kobakhidze and Lei Wu for providing valuable comments on this manuscript. This work was partially supported by the Australian Research Council. 

\appendix
\section{Appendix}\label{app:funcs}
Analogous to Eqn.~\ref{eqn:pheno_yt}, Higgs-fermion interactions may generally be parameterised as:
 \begin{align}
   \mathcal{L} &\supset -\frac{y_f}{\sqrt{2}} \overline{f}( g^S_{h\overline{f}f} +i\gamma^5 g^P_{h\overline{f}f}) f h,
\end{align}
where $y_f=\sqrt{2} m_f/v$, and in the SM $ g^S_{h\overline{f}f} =1$ and $ g^P_{h\overline{f}f} =0$. The diphoton decay rates for the scalar ($S$) and  pseudoscalar ($A$) Higgs are (see e.g. \cite{Spira:1997dg,Harlander:2005rq}):
\begin{align}
   \Gamma_S(h\rightarrow \gamma\gamma)&=\frac{m_h^3\alpha^2}{256\pi^3v^2} \bigg|\sum_{f} N_c Q_f^2 g^S_{h\overline{f}f} F^{1/2}_{s}(\tau_{h,f})+F^{1}(\tau_{h,W})\bigg|^2\\
      \Gamma_P(h\rightarrow \gamma\gamma)&=\frac{m_h^3\alpha^2}{256\pi^3v^2} \bigg|\sum_{f} N_c Q_f^2 g^P_{h\overline{f}f} F^{1/2}_{p}(\tau_{h,f})\bigg|^2,
\end{align}
where $\tau_{h,i}:= m_{h}^2/4m_i^2$, $Q_f$ is the charge of fermion $f$ in units of electric charge of positrons $N_C=1 (3) $ are the colour factors for leptons (quarks). The scaling function may be found, for example in \cite{Lee:2003nta}:
          \begin{align}
F^{1/2}_s(\tau) &=2\tau^{-1} [1+(1-\tau^{-1})f(\tau)]\\
F^{1/2}_p(\tau) &=2\tau^{-1} f(\tau)\\
F^{1}(\tau) &=-[2+3\tau^{-1}  +3\tau^{-1}(2 -\tau^{-1})f(\tau)]
          \end{align}   
          where $f(\tau)$ is in terms defined as: 
          \begin{align}
   \label{eqn:f_terms}
f(\tau) =-\frac{1}{2}\int^1_0 \ \frac{dy}{y} \ln [1-4\tau y (1-y)] = 
\begin{cases}
[\sin^{-1}(\sqrt{\tau})]^2, \quad & \tau\leq 1\\
-\frac{1}{4}\bigg[\ln\bigg(\frac{\sqrt{\tau}+\sqrt{\tau-1}}{ \sqrt{\tau}-\sqrt{\tau -1} }\bigg) -i\pi\bigg]^2, \quad& \tau\geq 1.
\end{cases}
         \end{align}
\input{bib.tex}

\end{document}

%% file: bib.tex
\bibliographystyle{unsrt}

%% file: main.bbl
\begin{thebibliography}{9}
\bibitem{CMS:2014ega}
  CMS Collaboration [CMS Collaboration],
and studies of the compatibility of its couplings with the standard model,''
  CMS-PAS-HIG-14-009.
\end{thebibliography}
